\documentclass[10pt,a4paper,times,oneside]{article}
\usepackage[margin=0.91in]{geometry}  

\usepackage{algorithm}
\usepackage{algorithmic}

\usepackage{amsmath}               
\usepackage{amssymb}
\usepackage{amsfonts}              
\usepackage{amsthm}                
\usepackage{caption}
\usepackage{color}
\usepackage{xcolor}
\usepackage{comment}
\usepackage{epsfig}
\usepackage{epstopdf}
\usepackage{graphicx, times}      
\usepackage{hyperref}
\usepackage{mathptmx}
\usepackage{mathalpha}
\usepackage{mathtools}
\usepackage{multicol}
\usepackage{multirow}
\usepackage{rotating}
\usepackage[numbers,sort&compress]{natbib}
\usepackage{pdflscape}
\usepackage{parskip}
\usepackage{subcaption}
\usepackage{tabularx} 
\usepackage{verbatim}
\usepackage{varwidth}

\newcommand*{\rom}[1]{\expandafter\@\romannumeral #1}

\newcommand{\bea}{\begin{eqnarray}}
	\newcommand{\eea}{\end{eqnarray}}
\newcommand{\bee}{\begin{eqnarray*}}
	\newcommand{\eee}{\end{eqnarray*}}

\begin{document}
\author{Romanshu Garg$^{1}$\footnote{romanshugarg18@gmail.com }, G. P. Singh$^{1}$\footnote{gpsingh@mth.vnit.ac.in}, Ashutosh Singh$^{2}$\footnote{ashuverse@gmail.com}
\vspace{.3cm}\\
${}^{1}$ Department of Mathematics,\\ Visvesvaraya National Institute of Technology, \\ Nagpur 440010, Maharashtra, India.
\vspace{.3cm}\\
${}^{2}$ Centre for Cosmology, Astrophysics and Space Science (CCASS),\\
GLA University, Mathura 281406, Uttar Pradesh, India.}
\date{}

\title{The $f(Q)$ gravity and affine EoS: Compatibility and observational constraints}

\maketitle

\begin{abstract}
We study the cosmological implications of barotropic fluid satisfying affine equation of state (EoS) in the General relativity and $f(Q)$ gravity framework. We describe the impact of affine EoS on the cosmic evolution in the model and derive the observational constraints on the model parameters. The models of General relativity and $f(Q)$ gravity may unify the scenario in which the universe transits from the decelerated expansion into the accelerated expansion. The model parameters are constrained by the Bayesian analysis based on $\chi^{2}$ minimization technique with the observational data of the Cosmic chronometer and Supernovae type Ia. The affine EoS model in the General Relativity possess quintessence kind of dark energy while it possess phantom kind of dark energy in the $f(Q)$ gravity. The present day values of the cosmological parameters along with the current age of the universe are compatible with the observations. We also probe the possibility of setting up the solution of General relativity model into the $f(Q)$ gravity.  
\end{abstract}
\section{Introduction}\label{sec:1}
The validation of accelerated cosmic expansion signifies the beginning of a new era in cosmological study. This phenomena was initially noticed while observing the type Ia supernovae \cite{1998AJ....116.1009R,1999ApJ...517..565P}. Subsequently, the universe's expansion has been confirmed by numerous observations \cite{knop2003new, tonry2003cosmological, 2020A&A...641A...6P}. An unusual fluid/field with negative pressure is cause of the universe's late-time accelerated expansion \cite{padmanabhan2003cosmological,carroll2001cosmological}. This kind of fluid/field has been termed as the `dark energy' whose exact nature remains mysterious \cite{padmanabhan2003cosmological,sahni2000case}. In other words, the dark energy, theoretically modeled to generate negative pressure has been believed to be responsible for the universe's accelerated expansion \cite{tsujikawa2010modified,nojiri2011unified,
nojiri2017modified,de2023finite,di2021realm,
heisenberg2024review}. The most widely accepted model of the universe is the $\Lambda$ cold dark matter ($\Lambda$CDM) model. In this model, the dark energy is described by the constant vacuum energy density which is visualized as the cosmological constant \cite{padmanabhan2003cosmological,carroll2001cosmological}. This model is consistent with the observations, however, it faces several challenges \cite{weinberg1989cosmological,padmanabhan2003cosmological,carroll2001cosmological,di2021realm}. The enormous difference between the theoretical and observational values of vacuum energy corresponding to cosmological constant causes the issue of fine tuning  \cite{padmanabhan2003cosmological,carroll2001cosmological}. In order to resolve the challenges of the $\Lambda$CDM model, the alternative approaches where the dark energy is a dynamical one is generally explored in cosmological modeling. \\
The issues related to the cosmic acceleration have been explored in dark energy models of the General relativity (GR) as well as the modified theories of gravity. Finding a modification of GR that allows cosmic acceleration without including any exotic components is an interesting option. One of the most well-known ideas to explain the dark contents of the universe is described by the $f(R)$ gravity \cite{buchdahl1970non}. This theory is a modification of General Relativity, where $R$ is the Ricci Scalar. Different theories of gravity have been explored to describe issues of dark energy and dark matter, see \cite{harko2011f,elizalde2010lambdacdm,
bamba2010finite,	harko2010f,nojiri2011unified,nojiri2017modified,
chaubey2016general,fqt,extcos,singh2020study,
as2022lyra,
Garg2023cfn,lalke2023late,as2024lyra,
capozziello2023role, singh2024abcde123,mgrav1} and references therein.\\
In the context of modified gravity theories, Jimenez et al. \cite{jimenez2018coincident} proposed a gravity theory based on the non-metricity os spacetime and is termed as $f(Q)$ gravity. This theory is based on an approach which considers a variation of the symmetric teleparallel equivalence to General Relativity, here $Q$ is a non-metricity scalar. The studies \cite{di2021realm,yang20212021}, suggest that the $f(Q)$ theory is one of the intriguing alternative gravity interpretations for explaining cosmological evolution. Lazkoz et al. \cite{lazkoz2019observational} investigated the accelerated expansion of the universe in  context of its consistency with observational data. The exact cosmic solutions that are both anisotropic and isotropic were studied by Esposito et al. \cite{esposito2022reconstructing}. Harko et al. \cite{harko2018coupling} developed a class of $f(Q)$ theories where $Q$ is non-minimally associated with the matter Lagrangian. In connection with the $f(Q)$ gravity theory, numerous further studies \cite{jimenez2018coincident,yang20212021, lazkoz2019observational, esposito2022reconstructing, harko2018coupling, maurya2023transit, pradhan2022quintessence,capozziello2024preserving,
nojiri2024well, hu2023nonpropagating,capozziello2022model,
khyllep2021cosmological,dimakis2021quantum,
dimakis2022flrw,heisenberg2024cosmological, capozziello2022comparing,koussour2023square,
rani2025physical, sarmah2024dynamical,nashed2024general, capozziello2024gravitational, parsaei2022wormhole,narawade2025stable,
Sahlu1,Sahlu2,singha2025,garg2024observational} suggest that the theory may have interesting cosmological and astrophysical implications, for review see \cite{heisenberg2024review}. \\
In this paper, we study the implications of dark fluid satisfying affine equation of state (EoS) in the General relativity and $f(Q)$ gravity. The issue of observational compatibility of model may imply for the observational constraints on model parameters. The dark fluid have been studied in the isotropic and anisotropic spacetime with the General relativity framework \cite{fluid1,fluid8,fluid9,fluid6,fluid3,fluid2,fluid4,
fluid5,fluid7,fluid11,fluid13,fluid14,fluid15,fluid16}. The affine EoS describes the hydro-dynamically stable as well as unstable fluid propagation. The speed of sound remains constant for the fluid. The stable dark energy evolution may be visualized with the fluid satisfying affine EoS given by $p_d=\alpha\rho_d-\rho_0$, where $\alpha$ and $\rho_0$ are constants. The classical stability criterion \cite{cs1,cs2} constrains $0\leq \alpha\leq 1$. In a non-conservative theory, this kind of EoS may naturally arise near the bouncing instant of the bouncing universe \cite{fluid14}. The low energy limits of this kind of EoS may have the $\Lambda$CDM limit \cite{fluid3}. Based on these peculiar properties, we aim to study the cosmological implications of this dark fluid in the General relativity and $f(Q)$ gravity. We search for an answer whether the General relativity solution of the affine EoS may have the $f(Q)$ gravity description or not? We also probe the nature of dark energy (phantom or quintessence or quintom) \cite{PhysRevLett.80.1582,caldwell2002phantom,cai2010quintom} in the affine EoS models of General relativity and $f(Q)$ gravity.  \\
This paper is organized in the following way: The affine EoS model of General relativity has been studied in Section (\ref{SecR1a}) to derive the Hubble parameter. An overview of the mathematical aspects of $f(Q)$ gravity in a flat FLRW background has been given in section (\ref{sec:2}) along with the cosmological solution of the affine EoS model in this gravity. In Section (\ref{sec:5}), the Bayesian statistical tools are used to investigate the compatibility  of cosmological solutions with the cosmic chronometer (CC) and supernovae type Ia (Pantheon) datasets. The cosmological dynamics based on the parameters such as the deceleration parameter, pressure, energy density, cosmographic parameter and the age of the universe are studied in section (\ref{sec:6}). In this section, we also probe whether solution of Section (\ref{SecR1a}) may be realized in the $f(Q)$ gravity framework or not? The conclusion with summary of results are given in Section (\ref{sec:7}).

\section{The affine EoS cosmological model in General Relativity}
\label{SecR1a}
The cosmic history of the universe is composed of the decelerating past and accelerated expansion during the present times. And, the universe is generally homogeneous and isotropic with flat spatial geometry to a great accuracy \cite{2020A&A...641A...6P}. This kind of universe expansion may be studied with the spatially flat FLRW spacetime, given by 
\begin{equation}{\label{6}}  
	ds^{2}=-dt^{2}+a^{2}(t) \left( dx^{2}+ dy^{2}+ dz^{2}\right).
\end{equation}
Here, the scale factor is $a(t)$ and, $H\equiv \frac{\dot{a}}{a}$ symbolizes the Hubble parameter with the overhead dot indicating the derivative with respect to cosmic time $t$. We proceed with $c=1$ unit in Eq. (\ref{6}). The Hubble parameter determines the expansion rate of universe. The field equations in the General relativity framework for FLRW metric (\ref{6}) are given by
\begin{equation}
	3H^2=k^2\rho, \quad 2\dot{H}+3H^2=-k^2p 
	\label{eq2}
\end{equation}
where $k^2=8\pi G$. To describe the homogeneous and isotropic universe, we proceed with the energy-momentum tensor ($T_{ab}$) of the perfect fluid as $ T_{ab} =  p g_{ab}+(p+\rho) u_a u_b$, where $a,b=0,1,2,3$. The perfect fluid's isotropic pressure and energy density are denoted as $p$ and $\rho$ respectively. The fluid's $4$-velocity $u_a$ satisfies the normalization condition $u_a u^a = -1$.\\
In order to explain the transiting universe evolution in this model, one may study the dark fluids which may unify the decelerating past with the accelerating present. In this model, we study the features of dark fluid satisfying affine equation of state (EoS). In general, the dark fluid may give rise to the transiting universe evolution (from matter dominated era into dark energy dominated era). We proceed by taking the fluid which may represent the hydro-dynamically stable dark energy. The EoS fluid has been assumed as \cite{fluid1,fluid2,fluid16}
\begin{equation}
	p_d=\alpha\rho_d-\rho_0, \label{eqr1}
\end{equation}
where $\alpha$ and $\rho_0$ are some constants. For $\rho_0=0$, one may have the usual EoS $p_d=\omega\rho_d$ for $\alpha=\omega$. The parameter $\alpha$ may be visualized as constant speed of sound and the classical stability criterion constrains $\alpha$ as $0\leq \alpha\leq 1$. The causality and stability issues are important as far as the physical acceptability of a cosmological model is concerned \cite{cs1,cs2}. In a given medium, the sound speed describes the propagation speed of the acoustic (pressure) waves. The behavior of squared sound speed may be described by ${c_s}^2=\frac{\partial{p_d}}{\partial{\rho_d}}$. The uncontrolled growth of energy density perturbations may be prevented by ${c_s}^2>0$. The prevention of causality violation would lead to ${c_s}^2<1$.\\
In this General relativity model, the explicit assumption is that the energy density and pressure appearing in the Eq. (\ref{eq2}) is given by $\rho_d$ and $p_d$ respectively, satisfying affine EoS (\ref{eqr1}). This kind of assumption for present universe will be purely valid during the era governed by matter and dark energy.\\ 
The energy conservation equation for the dark fluid may be written as
\begin{equation}
	\dot{\rho}_d+3H(\rho_d+p_d)=0.
	\label{eqr2}
\end{equation} 
One may observe that for $\rho_d< \frac{\rho_0}{1+\alpha}$, one may have the phantom evolution of dark energy in model. For $\rho_d= \frac{\rho_0}{1+\alpha}$, the dynamics of constant vacuum energy may be visualized. For $\rho_d> \frac{\rho_0}{1+\alpha}$, the null energy condition will be satisfied in model, as a consequence the fluid distribution may either have the quintessence domination or the normal matter (composed of matter or radiation) domination. We take the usual scale factor - redshift relation $\frac{a_0}{a}=1+z$, where $a_0$ is the present day scale factor value. By convention $a_0=1$.  \\
The dark fluid conservation equation (\ref{eqr2}) may lead to the $\rho_d$ form depending on the $z$ as 
\begin{equation}{\label{502}}
	\rho_{d}=\frac{\rho_0 + (1+z)^{3(\alpha +1)}}{\alpha +1}.
\end{equation}
In this case, the Hubble parameter describing the expansion rate of universe will be 
\begin{equation}{\label{301}}
	H(z)=H_{0}\sqrt{  \frac{(1+z)^{3(1+\alpha)} +\rho_{0}}{\rho_{0}+1}}, 
	\quad \text{where,} \quad H_0=\sqrt{\frac{\rho_0 +1}{3(1+\alpha)}}.
\end{equation}
where $H_0$ describes the present day Hubble parameter. One may note that $H(z=0)=H_0$ is being identically satisfied from this expression. The above Hubble parameter may describe the cosmic evolution governed by the fluid following affine equation of state (\ref{eqr1}). The observational constraints on the model parameters $\alpha,\rho_0$ and $H_0$ may describe the observational compatibility of the model along with the validity of causality principle in this model. We have described - in detail - the observational compatibility of this model in Section \ref{sec:5}.

\section{An overview of $ f(Q)$ gravity and its field equations}
\label{sec:2}
The $f(Q)$ gravity is based on the non-metricity scalar $Q$. The $f(Q)$ gravity is characterized by the following action \cite{jimenez2018coincident}
\begin{equation}{\label{1}}
S = \int \sqrt{-g}  \, d^4x \left[-\frac{1}{2}f(Q) + \mathcal{L}_m\right]. 
\end{equation}
Here, $f(Q)$ represents an arbitrary function of the non-metricity scalar $Q$. The symbol $g$ denotes metric tensor's determinant $g_{ab}$ and the Lagrangian density for matter is symbolized by $\mathcal{L}_m$. The quantity $Q$ can be characterized by two distinct traces as
\begin{equation}{\label{2}}
Q_{\sigma} = Q_{\sigma a}{}^{a}, \qquad \tilde{Q}_{\sigma} = Q^{a}{}_{\sigma a}.
\end{equation}
Furthermore, the non-metricity scalar $(Q)$ may be expressed by the contraction of above tensors and may be defined as:
\begin{equation}{\label{3}}
Q=-Q_{\sigma ab} P^{\sigma ab}.
\end{equation}
Additionally, the super-potential tensor (which is a conjugate of non-metricity) can be introduced as:
\begin{equation}{\label{4}}
4P^\sigma_{\phantom{\sigma}ab} = 2Q_{(a \phantom{\sigma}b )}^{\phantom{\mu}\sigma} - \tilde{Q}^\sigma g_{ab} -Q^\sigma_{\phantom{\sigma}ab}   - Q^\sigma g_{ab}  - \delta^\sigma_{(a} Q_{b)}^{\phantom{\nu)}}.
\end{equation}
The field equation may obtained by varying the action (\ref{1}) with respect to the metric and are expressed as \cite{jimenez2018coincident}
\begin{equation}{\label{5}}
\frac{2}{\sqrt{-g}} \nabla_\sigma \left( \sqrt{-g} f_Q P^{\sigma}_{ab} \right) + \frac{1}{2} g_{ab}f + f_Q \left( P_{a\sigma\beta} Q^{\sigma\beta}_{\phantom{\sigma\beta}b} - 2Q_{\sigma\beta a} P^{\sigma\beta}_{\phantom{\sigma\beta}b} \right) = T_{ab},
\end{equation}
where, $f_Q \equiv \frac{df(Q)}{dQ}$ and, the energy-momentum tensor may be written as $
T_{ab}=-\frac{2}{\sqrt{-g}} \frac{\delta(\sqrt{-g} \mathcal{L}_m)}{\delta g^{ab}}$. Here, we take the unit $8\pi G = c = 1$. Under these conditions, the non-metricity scalar may be described as $Q=6H^{2}$ \cite{jimenez2018coincident}. For the metric (\ref{6}), the Friedmann equations can be expressed as \cite{jimenez2020cosmology}: 
 \begin{eqnarray}
H^2 = \frac{1}{6f_Q} \left( \rho + \frac{f}{2} \right), \quad \dot{H} +  3H^2 + \frac{H \dot{f}_Q}{f_Q}  = \frac{1}{2f_Q} \left( \frac{f}{2}-p  \right).\label{9}
\end{eqnarray}
If $f(Q)$ is assumed to be $f(Q)=Q$ \cite{jimenez2020cosmology}, one may obtain the standard field equations of the General Relativity. By considering $f(Q)=Q+F(Q)$, the above field equations (\ref{9}) may also be expressed as 
\begin{equation}\label{10}
3H^2 = \rho + \frac{F}{2} - Q F_{Q}, \quad \dot{H}(2QF_{QQ}+F_{Q}+ 1)+\frac{1}{4} \left(2QF_Q - F + Q \right) = -\frac{p}{2}.
\end{equation}
The Eq. (\ref{10}) may be seen as modifications of the General Relativity equations, incorporating an additional term that arises from the non-metricity of space-time. 
This term may exhibit a characteristics similar to the dark energy of universe, i.e. $p_{Q}=p_{d}$ and $\rho_{Q}=\rho_{d}$. The pressure and density contributions to dark energy may be represented by $p_{d}$ and $\rho_{d}$ respectively, which may be caused by the non-metricity of spacetime. We define 
\begin{equation}{\label{14}}
\rho_{d} = \frac{F}{2} -QF_{Q}, \quad p_{d} = 2\dot{H}(2QF_{QQ} + F_Q) - \rho_{d}.
\end{equation}
The decelerating universe expansion occurs due to dominance of cold dark matter having energy density $\rho_m$ and pressure $p_m=0$. The dark energy is an exotic component having negative pressure and it is a widely accepted fact that the dark energy drive the accelerated expansion of the universe. In the context of $f(Q)$ gravity having field equations (\ref{10}), we consider the universe composed of matter (having energy density $\rho_m\equiv \rho,p_m=0$) and dark energy density $\rho_d$ governed by Eq. \ref{eqr2}. These assumptions are motivated by the fact that matter (cold dark matter and baryonic matter) and dark energy are the dominant components of the present day universe. In this model, we probe for the universe's evolution in $f(Q)$ gravity when the matter (following $p_m=0$) and dark energy (following Eq. (\ref{eqr1})) are the dominant components of the universe. \\     
For the present case, the energy density of matter is $\rho_m$ satisfying EoS $p_m=0$. The conservation equation of matter $\dot{\rho}_m+3H\rho_m=0$ would lead to $\rho_m=\rho_{m0}a^{-3}$, where $\rho_{m0}$ is energy density of matter at $z=0$. And, $\rho_d$ is the dark energy density following Eq. (\ref{eqr2}). In this case, the Hubble parameter may be obtained as 
\begin{equation}{\label{302}}
H(z)=H_{0}\sqrt{\Omega_{m0}   (1+z)^3 + \Omega_{\Lambda0}+ \Omega_{a0}(1+z)^{3(1+\alpha)}}
\end{equation}
where $\Omega_{m0}\equiv \frac{ \rho_{m0}}{3H_{0}^2}$, $\Omega_{\Lambda0}\equiv \frac{ \rho_{0}}{3(1+\alpha)H_{0}^2}$ and $\Omega_{a0}\equiv \frac{ \rho_{\alpha0}}{3(1+\alpha)H_{0}^2}$. At $z=0$ in present era, $H=H_0$ with constraint $\Omega_{m0}+\Omega_{\Lambda 0}+\Omega_{a 0}=1$. This Hubble parameter may describe the cosmic dynamics in $f(Q)$ gravity where the fluids are composed of matter and dark energy satisfying $p_m=0$ and $p_d=\alpha\rho_d-\rho_0$ respectively. Note that we have not assumed any form of $f(Q)$ in this modified gravity. \\
One may obtain the form of $f(Q)$ by solving the equation 
\begin{equation}{\label{1011}}
\frac{F(Q)}{2} -Q\frac{\partial{F(Q)}}{\partial Q} = \rho_{d}(z)
\end{equation}
where $\rho_{d}$ can be taken from Eq. (\ref{502}). We solve the equation (\ref{1011}) numerically with numerical solver with the initial condition $f(0)=6H_{0}^{2}(2+\Omega_{d0})$, where we take $\rho_0$ to be unity. In this study, we ignore the contribution of radiation since it is observationally very small in the late-time universe \cite{2020A&A...641A...6P}. We probe the observational compatibility of these models with the observational data and derive the constraint on model parameters. We further describe the corresponding differences in universe evolution for these cases of Section \ref{SecR1a} and \ref{sec:2}.
\begin{figure}[!htb]
\captionsetup{skip=0.3\baselineskip,size=footnotesize}
   \begin{minipage}{0.5\textwidth}
     \centering
     \includegraphics[width=7.5 cm,height=6.6cm]{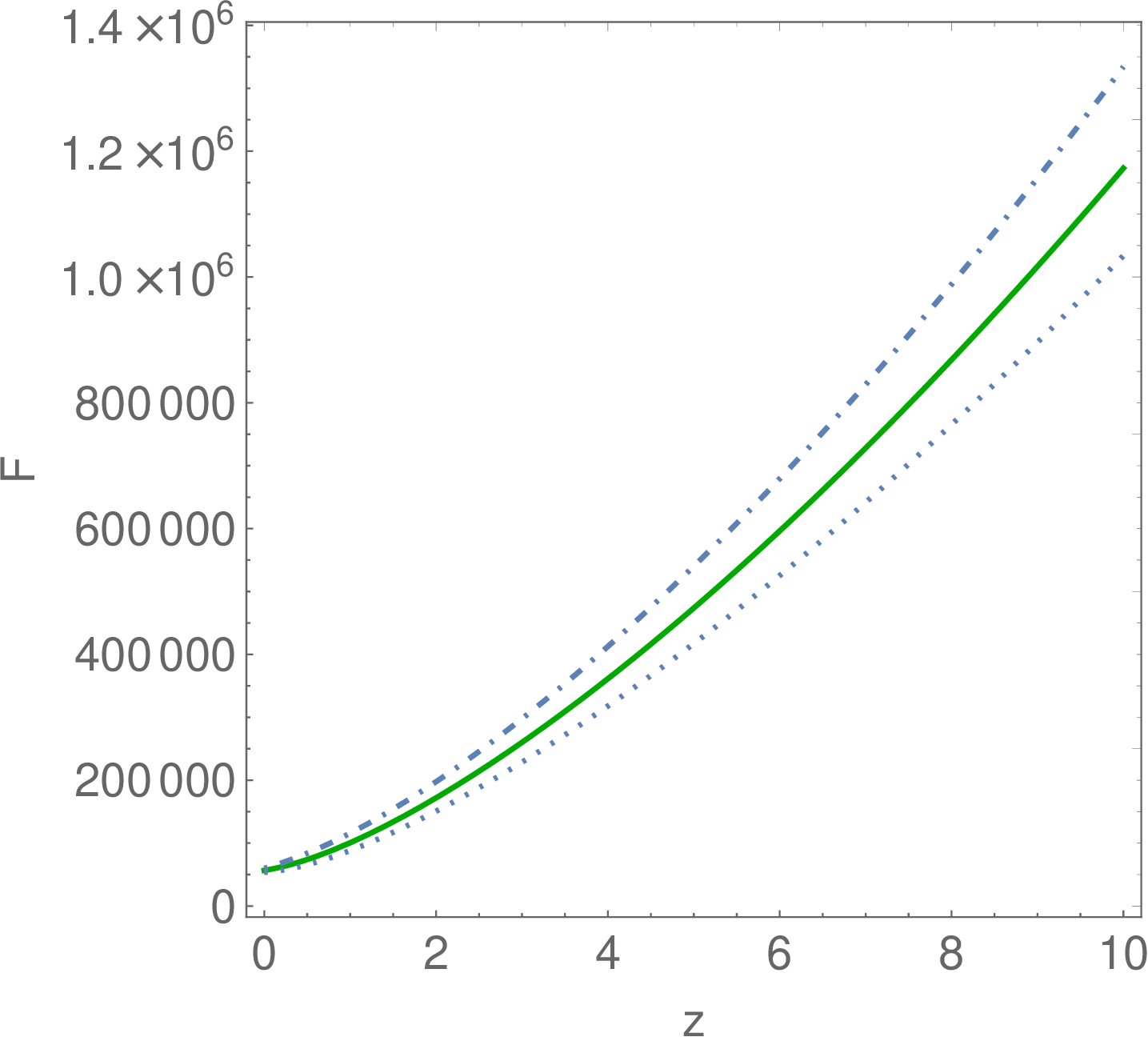 }
\caption{$F$ versus ${z}$, where solid and dotted curves represent the curve corresponding to central and $1\sigma$ values.}
\label{fig:109}
    \end{minipage}\hfill
   \begin{minipage}{0.5\textwidth}
     \centering
     \includegraphics[width=7.5 cm,height=6.6cm]{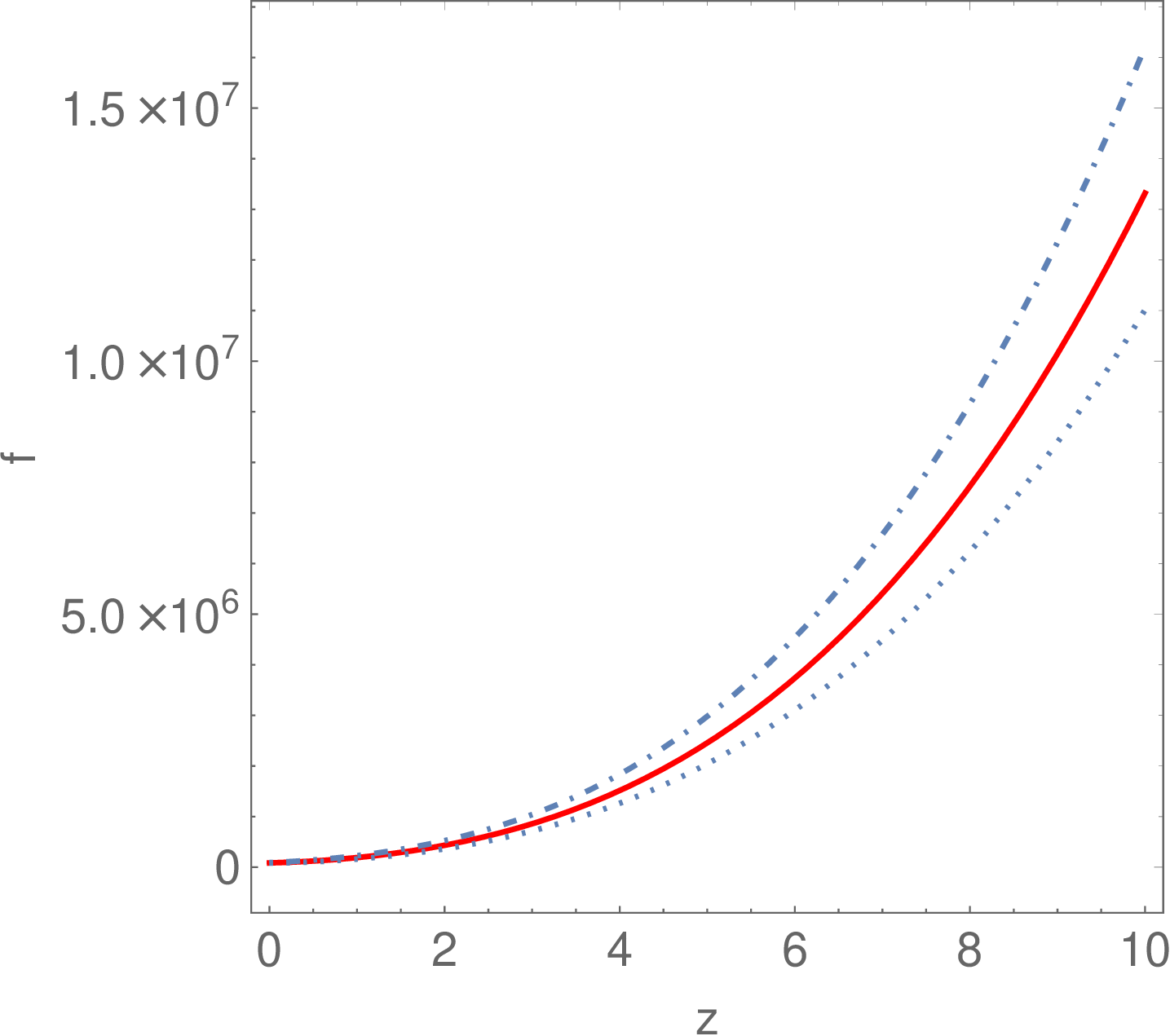}
  \caption{$f$ versus $z$, where solid and dotted curves represent the curve corresponding to central and $1\sigma$ values. }
\label{fig:113}
   \end{minipage}
\end{figure}
The evolution of functions $F(z)$ and the full $f(Q)=Q+F(z)$ with redshift is presented in figures $(\ref{fig:109})$ and $(\ref{fig:113})$ respectively. The figures are illustrating the monotonically decreasing behavior of $f$ and $F$ from the past (high $z$) to the present epoch ($z=0$). This decreasing trend is a direct consequence of the cosmic expansion. In the past, the higher energy density of the universe corresponded to a higher non-metricity scalar $Q$ and thus a larger value of the function $f(Q)$. As the universe expands and the energy density drops, the value of non-metricity function $F(Q)$ diminishes. This behavior aligning with observed cosmological evolution.
\section{Observational constraints and results}
\label{sec:5}
In this section, we focus on describing the data sets and methodology that are utilized to analyze the models. We nomenclate the General relativity model with affine EoS described by $H(z)$ in Eq. (\ref{301}) as Model I and the $f(Q)$ gravity model (\ref{302}) as Model II. We also describe the constraints on model parameters of these models. In other words, the consistency of the Hubble parameter (\ref{301}) and (\ref{302}) with the data sets given by the Cosmic Chronometer (CC) and combined data (consisting of Pantheon and CC) are used in the MCMC analysis. We constrain the parameters using $\chi^{2}$ minimization method and the MCMC method is implemented with the emcee Python library \cite{foreman2013emcee}. 
\subsection{The Cosmic chronometer data}
\label{sec:5.1}
\begin{figure}
	\begin{center}
\includegraphics[width=12cm, height=7cm]{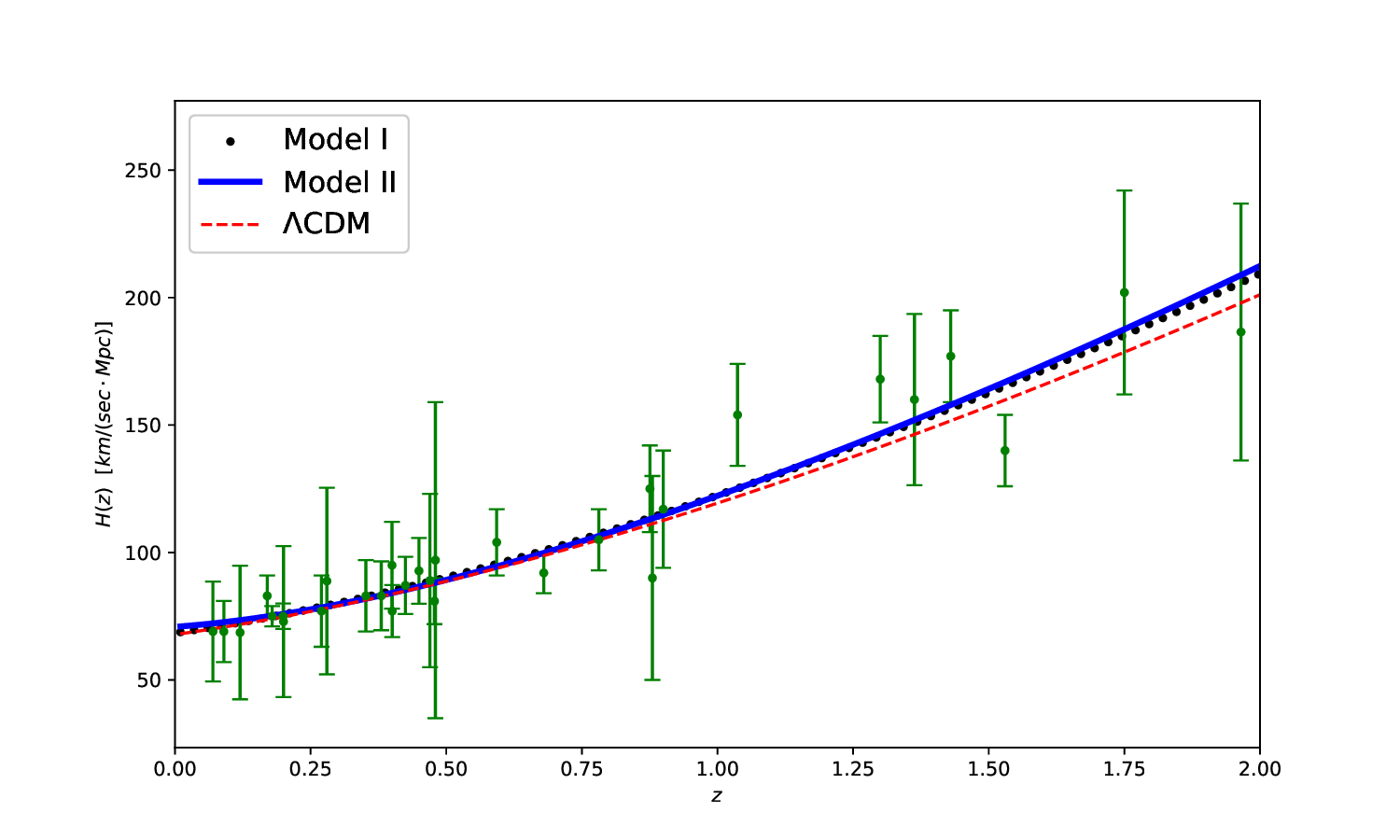}
\caption{The best fit Hubble parameter curves with $z$ as compared to the $\Lambda CDM$ model.}
\label{fig:1}
\end{center}
\end{figure}

The model parameters are constrained using the Cosmic Chronometer (CC) data consisting of $31$ data points in the redshift range of $0.07 \leq z \leq 1.965$. These data are collected using the differential ages of galaxies approach \cite{vagnozzi2021eppur,GSSharov}. The cosmic chronometer observations are based on the principle introduced by Jimenez and Loeb \cite{jimenez2002constraining}. It describes a relationship between redshift $(z)$, cosmic time $t$, and the Hubble parameter $H(z)$ as $H(z)=\frac{-1}{(1+z)}\frac{dz}{dt}$. By minimizing the $\chi^{2}$ function, we determine the model parameters' central values  with $1\sigma$ errors in the emcee \cite{foreman2013emcee} package. The $\chi^{2}$ function has been taken as \cite{fluid16, lalke2024cosmic,mandal2024late,aspdu,sma2026}
\begin{equation}{\label{24}}
\chi^{2}_{CC}(\theta)=\sum_{i=1}^{31} \frac{[H_{th}(\theta,z_{i})-H_{obs}(z_{i})]^{2}    }{ \sigma^{2}_{i}}.   
\end{equation} 
Here, $H_{th}$ is used to indicate the theoretical values of the Hubble parameter, the observed values are indicated by $H_{obs}(z_i)$ and the notation $\sigma_i$ indicates the standard deviation for each $H_{obs}(z_i)$ observed value. The emcee \cite{foreman2013emcee} results yield the median values of posterior distributions for the model parameters. Tables (\ref{table:1}) and (\ref{table:2}) provide summaries of these values for Model I and Model II, respectively. The error bars of the CC points alongside the best fit Hubble parameter curve derived from equation (\ref{301}), (\ref{302}) are illustrated in figure $(\ref{fig:1})$.
It may be observed that the cosmic dynamics of the Model I and II are deviating from the $\Lambda$CDM model in the past where $z>1$. The transition redshift however remains close to the $\Lambda$CDM model. It is worthwhile to observe from these results that the Model I may describe the transiting universe evolution from the decelerated expansion into the accelerated expansion. And, the cosmic dynamics in Model II will remains similar to the Model I. 
\subsection{The Pantheon data}
\label{sec:5.2}
\begin{figure}[ht!]
\centering
\includegraphics[scale=0.6]{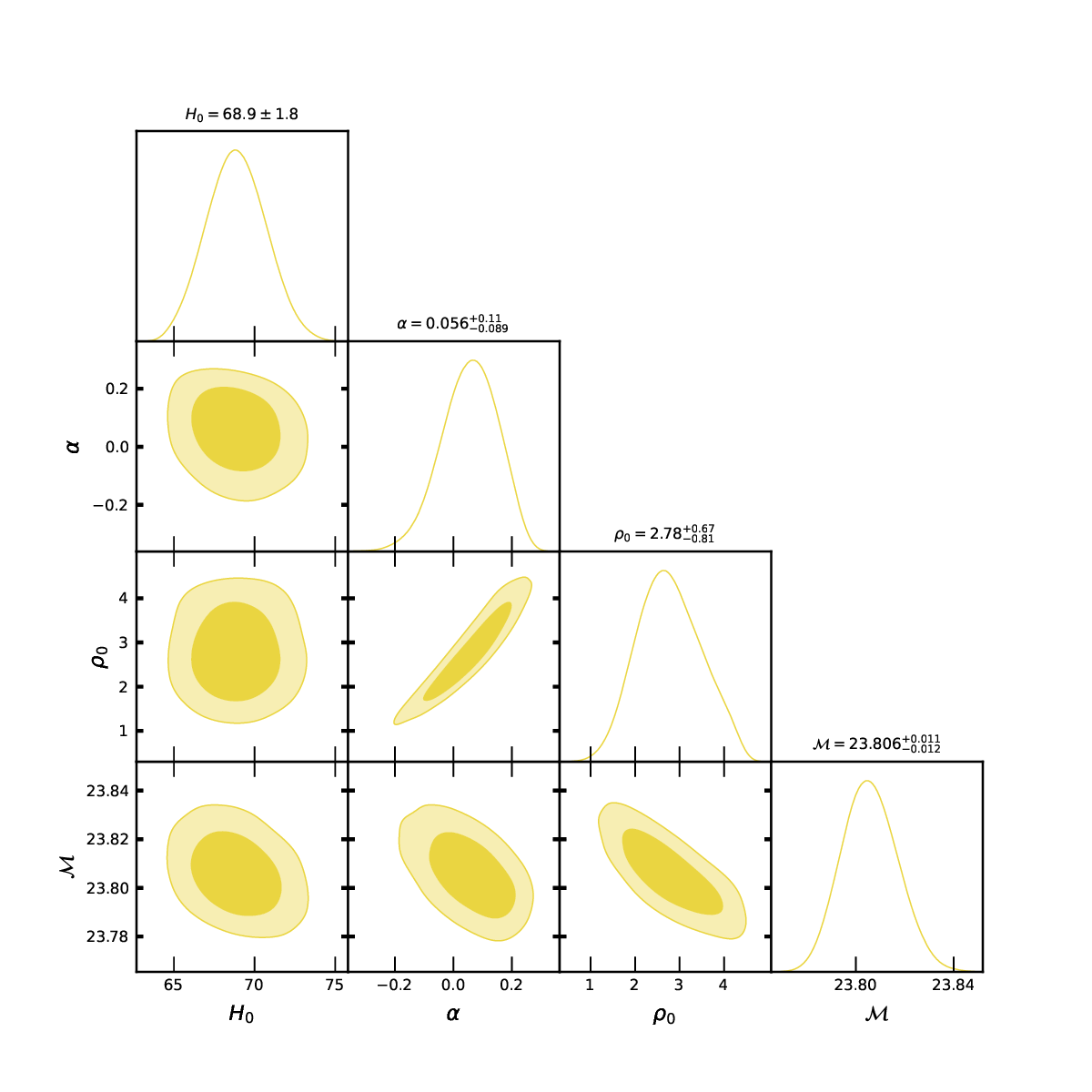}
\caption{Model I: Marginalized $1$D and $2$D posterior contour map with median values of parameters using Joint data set.}
\protect\label{fig:5}
\centering
\end{figure}

\begin{figure}[ht!]
\centering
\includegraphics[scale=0.6]{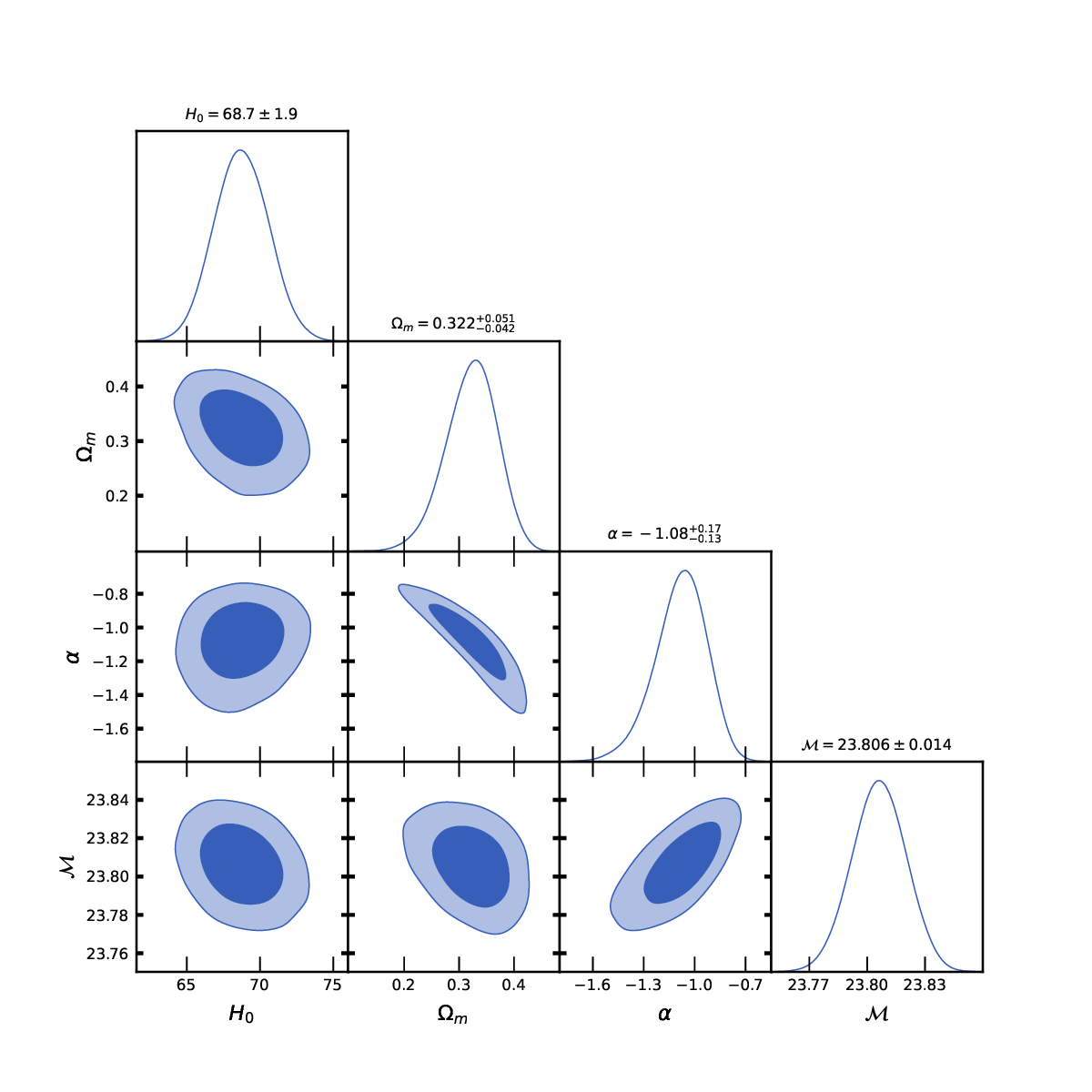}
\caption{Model II: Marginalized $1$D and $2$D posterior contour map with median values of parameters using Joint data set.}
\protect\label{fig:100}
\centering
\end{figure}

The Pantheon data is composed of supernovae type Ia (SNIa) observations identified in the low and high redshift ranges \cite{scolnic2018complete}. This dataset includes $1048$ data points within the redshift interval $0.01 \leq z \leq 2.26$. For the purpose of examining cosmic expansion of the observable universe, SNIa are widely adopted as the standard candles with their theoretical apparent magnitude $m_{th}(z)$ are given by:
\begin{equation}{\label{7a}}
m_{th}(z)=M+5\log_{10}\left[\frac{d_{L}(z)}{\text{Mpc}}\right]+25,
\end{equation}
where, $M$ is absolute magnitude, $c$ denotes the speed of light, $z$ represents the SNIa redshift in the CMB rest frame and $d_{L}(z)=c(1+z)\int_{0}^{z}\frac{dz'}{H(z')}$ is the luminosity distance. The luminosity distance $d_L(z)$ is typically replaced with its Hubble-free counter part $(D_{L}(z) \equiv H_{0}d_{L}(z)/c)$. We can also rewrite equation (\ref{7a}) in the following way:
\begin{equation}{\label{9a}}
m_{th}(z)=M+5\log_{10}\left[D_{L}(z)\right]+5\log_{10}\left[\frac{c/H_{0}}{  \text{Mpc}}\right]+25. 
\end{equation}
In the $\Lambda$CDM model framework, a degeneracy will exist between $H_{0}$ and $M$ \cite{ellis2012relativistic,asvesta2022observational}. We take $\mathcal{M}$ as a combination of these parameters \cite{asvesta2022observational}:
\begin{equation}{\label{10a}}
\mathcal{M}\equiv M+5\log_{10} \left[\frac{c/H_{0}}{ \text{Mpc}}\right]+25
\end{equation}
where $H_{0}=h \times 100 \ Km/s/Mpc$ and thus one can write $\mathcal{M} =42.39+M-5\log_{10}(h)$. For the Pantheon data, this parameter is used along with the relevant $\chi^{2}$ in the MCMC analysis defined 
as \cite{asvesta2022observational,mandal2024late,
lalke2024cosmic,fluid16,aspdu}
\begin{equation}{\label{11a}}
\chi^{2}_{P}= \nabla m_{i}C^{-1}_{ij}\nabla m_{j},
\end{equation} 
where, $\nabla m_{i}=m_{obs}(z_{i})-m_{th}(z_{i})$, the theoretical apparent magnitude $m_{th}(z_i)$ is calculated using equation (\ref{9a}) and the symbol $m_{obs}(z_i)$ is used to denote the apparent magnitude of SNeIa observations at redshift $z_i$ and $C_{ij}^{-1}$ is covariance matrix's inverse. The luminosity distance is influenced by the Hubble parameter. Following this approach, we use equations (\ref{301}) and (\ref{302}) and the emcee package \cite{foreman2013emcee} to determine the maximum likelihood estimate using the joint (CC+Pantheon) dataset. For computing the maximum likelihood estimate using the joint data, we define the joint $\chi^{2}$ as $\chi^{2}_{CC}+\chi^{2}_{P}$. \\
For Model I and II, the derived posterior contour maps with $1\sigma$ and $2\sigma$ confidence levels along with the one dimensional marginalized distribution are given in Fig. (\ref{fig:5}) and (\ref{fig:100}). The summary of cosmological parameters in Table $(\ref{table:1})$ and $(\ref{table:2})$ suggest that the models are compatible with observations. In order to obtain the posterior distribution of model parameters, we use the Hubble parameter (\ref{301}) for the Model I. In the Model II, we use the Hubble parameter (\ref{302}) with the constraint $\Omega_{m0}+\Omega_{\Lambda 0}+\Omega_{a0}=1$. In the MCMC analysis with CC and joint data, we observe that the posterior distributions subjected to parameter $\Omega_{a0}$ and $\alpha$ possesses the degeneracy. In other words, the parameter space remain unbounded and it lead to the ill-defined, un-physical bounds on parameter $\alpha$ and $\Omega_{a0}$. Further, we fix the parameter $\Omega_{\Lambda 0}=0.01$ with the assumption $\Omega_{\Lambda 0}<\ll \Omega_{a0}$. For this case, the contour map with $1\sigma$ and $2\sigma$ limits are shown in Fig. (\ref{fig:100}). In this case, we observe that the parameter $\alpha$ is negative within $1\sigma$. It simply mean that this  model is not classically stable. In the case without assumption $\Omega_{\Lambda 0}<\ll \Omega_{a0}$ also, the $\alpha$ remains negative leading to the classically unstable universe evolution in model II. In contrast, we observe in Model I that the universe evolution will be classically stable subjected to the constraint on $\alpha$. For the Model I, within $1\sigma$ limit on $\alpha$, one may have $(-0.022,0.086)$ based on the CC constraints and $(-0.033,0.166)$ based on joint constraints. These constraints are suggesting that there is a broader region where $\alpha$ is positive and, thus will be classically stable in the region where $0\leq \alpha\leq 1$. It simply mean that there are fewer $\alpha$ values for which Model I may be unstable. Hence, this model is not fully unstable.  In the next section, we describe in detail on the cosmic dynamics possessed by the models. \\
For Model I: For the CC data set, we use $48$ random walkers and $60000$ iterations (steps) for the emcee based MCMC analysis. We select uniform priors on $H_{0}, \alpha$ and $\rho_{0}$ for the CC data set. We take the range $60 < H_{0} < 80, -1 < \alpha <1 $ and $  0.0 < \rho_{0} < 4.5$. For the joint data set, we use $48$ random chains (walkers) and $16000$ iterations (steps) for the  emcee based MCMC analysis. We select uniform priors on $H_{0}, \alpha, \rho_{0}$ and $\mathcal{M}$ for the joint data set. We take the range $60 < H_{0} < 80, -1 < \alpha < 1, 0.0 < \rho_{0} <4.5 $ and $ 23.5 < \mathcal{M} < 23.85 $. \\
For Model II: For the CC data set, we use $48$ random walkers and $60000$ iterations (steps) for the  emcee based MCMC analysis. We select uniform priors on $H_{0}, \Omega_{m}$ and $\alpha$ for the CC data set. We take the range $60 < H_{0} < 85, 0.01 < \Omega_{m} <0.6 $ and $  -3 < \alpha < 1.0$. For the joint data set, we use $48$ random walkers and $16000$ iterations (steps) for the  emcee based MCMC analysis. We select uniform priors on $H_{0}, \Omega_{m}$ and $\alpha$ and $\mathcal{M}$ for the joint data set. We take the range $60 < H_{0} < 85, 0.01 < \Omega_{m} <0.6, -3 < \alpha < 1.0$ and $ 23.5 < \mathcal{M} < 24 $.
\\
\begin{table}[htbp]
\footnotesize
\centering
\begin{tabular}{|c|c|c|c|c|c|c|c|c|c|c|}
\hline
Dataset & $H_{0}$  & $\alpha$  & $\rho_0$ & $\mathcal{M}$ & $q_{0}$ & $z_{t}$ & $t_{0}$ (Gyr) & $j_{0}$ & $\omega_{e}(z=0)$ \\
\hline
CC & $68.6^{+1.0}_{-1.0}$ & $0.032^{+0.054}_{-0.054}$ &$2.52^{+0.51}_{-0.70}$ & - & -0.560 & 0.638 & $13.51^{+0.08}_{-0.37}$ & $1.145$ & $-0.7068$\\
\hline
joint  & $68.9^{+1.8}_{-1.8}$  & $0.056^{+0.11}_{-0.089}$  &  $2.78^{+0.67}_{-0.81}$ & $23.806^{+0.011}_{-0.012}$ & -0.580 & 0.65 & $13.41^{+0.65}_{-1.04}$ & $1.255$ & $-0.7206$\\
\hline
\end{tabular}
\caption{ {\bf{Model I}}: 
The constrained values with $q_{0}, z_{t}$, $t_{0}$ $j_0$, where $H_0$ is in $Km/s/Mpc$.}
\label{table:1}
\end{table}
\normalsize
\begin{table}[htbp]
\footnotesize
\centering
\begin{tabular}{|c|c|c|c|c|c|c|c|c|c|}
\hline
Dataset & $H_{0}$ & $\Omega_m$  & $\alpha$ &  $\mathcal{M}$  &  $q_0$ & $z_t$ & $j_0$ & $t_{0}$ (Gyr) & $\omega_e(z=0)$  \\
\hline
CC & $71.0^{+5.5}_{-8.2}$ & $0.305^{+0.067}_{-0.056}$ &$-1.33^{+0.73}_{-0.49}$ & - & $-0.881$ & $0.62$ & $2.352$ & $13.76$ & $-0.921$ \\
\hline
joint  & $68.7^{+1.9}_{-1.9}$  & $0.322^{+0.051}_{-0.042}$  &  $-1.08^{+0.17}_{-0.13}$ &  $23.806\pm 0.014$ & $-0.597$ & $0.63$ & $1.259$ & $13.59$  & $-0.7314$ \\
\hline
\end{tabular}
\caption{ {\bf{Model II}}: The constrained values with $q_{0}, z_{t}$, $t_{0}$ and $j_0$, where $H_0$ is in $Km/s/Mpc$.}
\label{table:2}
\end{table}
\normalsize

\section{Cosmic dynamics and the physical behavior}
\label{sec:6}
In this section, we investigate the dynamical characteristics of the universe evolution in model I and II by using the observationally constrained values of model parameters. 
\subsection{Cosmographic parameter}\label{sec:6.4}
The study of cosmography was initially introduced by Weinberg \cite{weinberg2008cosmology}, who used a Taylor series to introduce the scale factor that rose around present time $t_{0}$. A detailed overview of cosmography may be reviewed for different aspects of cosmographic parameters \cite{extcos}. The Hubble parameter ($H$), has been considered as a variable observable quantity. The behavior of the Hubble parameter (Fig. \ref{fig:1}) highlights the expansion rate of the universe in considered models. The evolution of deceleration parameter $q$ is characterized through the second-order derivative of scale factor ($a$) \cite{mukherjee2016parametric}. To understand the universe's cosmic evolution, the snap ($s$) and jerk ($j$) parameters serve as important tools. One of the cosmological parameters is the deceleration parameter that helps to characterize the universe's expansion behavior. It is basically used to track down the rate of slowing down of the universe's expansion. Mathematically, using the Hubble parameter and its derivative, it may be expressed as 
\begin{equation}{\label{29}}
	q = -1 - \frac{\dot{H}}{H^2}.
\end{equation}
This parameter serves as a tool to assess whether the universe is expanding at an increasing or decreasing rate. Depending on the value of $q$, different phases of cosmic expansion may be visualized. If $q$ is greater than zero $(q>0)$, the universe is in a decelerated phase. For $q<0$, the universe will expand at an accelerating rate. For $-1<q<0$ the expansion follows a power-law acceleration. A super-exponential expansion era occurs in the universe for $q<-1$, while a de Sitter expansion when $q=-1$ \cite{mandal2022epjp,SINGH2024865,asgrg2}. For matter and radiation-dominated universe expansion, the deceleration parameter will be $q=\frac{1}{2}$ and $q=1$ respectively. By using (\ref{301}), (\ref{302}) and (\ref{29}), we may obtain for Model I and II as 
\begin{eqnarray}
		q(z)=-1+ \left( {\frac{3(1+\alpha)(1+z)^{3+3\alpha}}{  2(\rho_0+(1+z)^{3+3\alpha})}} \right), \label{30} \ \  \text{and} \\
	  	q(z)=-1+\frac{3 (z+1)^3 ((\alpha +1) \Omega_{a0} (z+1)^{3 \alpha }+\Omega_{m0})}{2 ((z+1)^3 (\Omega_{a0} (z+1)^{3 \alpha }+\Omega_{m0})+\Omega_{\Lambda 0})},\label{104} 
\end{eqnarray}
respectively.
	\begin{figure}[ht!]
\centering
\includegraphics[scale=0.6]{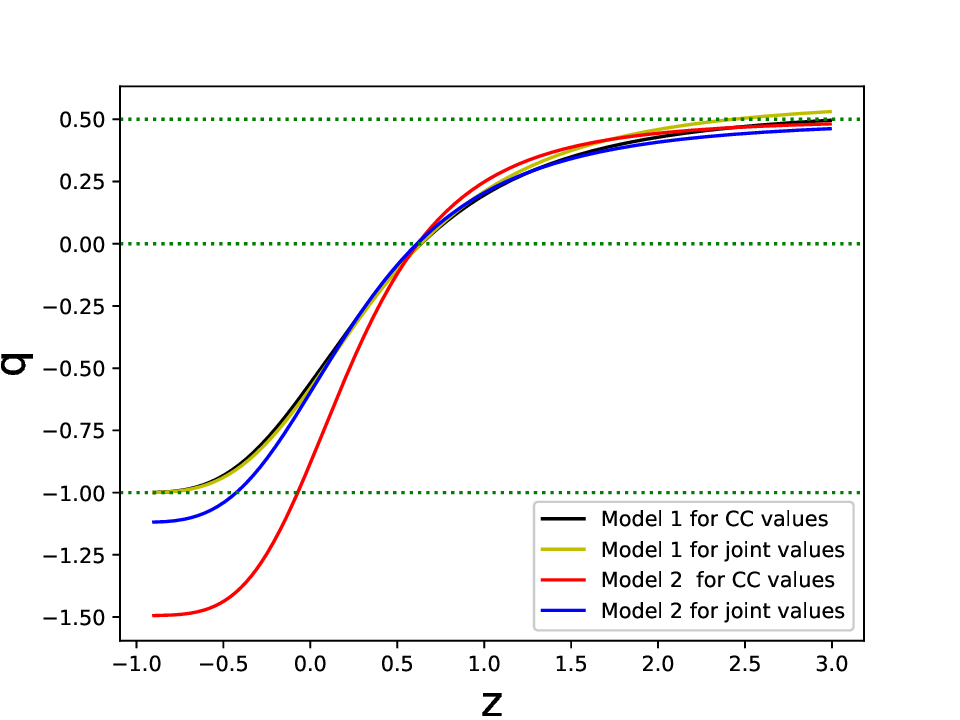}
\caption{$q(z)$ versus $z$.}
\protect\label{fig:7}
\centering
\end{figure}
The behavior of $q(z)$ of model I (\ref{30}) and model II (\ref{104}) for the median values of parameters are illustrated in Fig. $(\ref{fig:7})$. It indicates that in the early universe, $q(z)$ is positive, while in the present times, it has becomes negative ($q<0$) due to the domination of dark energy in these models.\\
The deceleration parameter of these models provides insight into the universe's transition from deceleration into acceleration of expansion. In model I, the present day value of deceleration parameter is $q_0=-0.56$ and $q_0=-0.58$ for CC and joint data respectively, and the value of transit redshift are $z_t=0.638$ (for CC data) and $z_t=0.65$ (for joint data). The redshift value at which $q=0$ is termed as the transit redshift, since the behavior of $q$ is changing from positive to negative in the cosmic evolution. The accelerated universe expansion may be visualized in the model I for $z<0.638$. For Model II, the transition from decelerated to accelerated expansion occurs at $z_t=0.62$ based on CC estimates and at $z_t=0.63$ using joint estimates. For CC estimates, the deceleration parameter's current value  is $q_0= -0.881$, while for joint estimates, it is $q_0 = -0.597$. These negative values of deceleration parameters of  model I and II are confirming that the universe is expanding with acceleration in the present era $(z=0)$. And, thus it has been observed that the Hubble parameter may yield the transiting universe evolution in model I governed by affine EoS. In the Model II, the dark energy introduced by the non-metricity may yield the accelerated expansion at late-times while the cold dark matter yields the decelerated universe expansion at early times in model.\\
The behavior of $q$ in Fig. (\ref{fig:7}) suggests that the model I will possess the matter-dominated era of universe expansion which may be confirmed with $(q\rightarrow \frac{1}{2})$ during early times. An accelerated phase $(q < 0)$ in the present era and ultimately converges to the de Sitter scenario $(q=-1)$ as $z\rightarrow -1$. This model may explain the accelerated expansion universe at the present times, which is consistent with the observational data. It has been observed that the deceleration parameter of Model II is consistent with the current phase of accelerated expansion. In Fig. (\ref{fig:7}), the accelerated universe expansion of present times may reach to super-exponential expansion having $q<-1$ in the future. It is an important implication of the Model II, which is classically unstable due to $\alpha<0$. \\ 
 The jerk parameter $(j)$ indicates the rate at which the universe's acceleration or deceleration changes over time. The snap parameter $(s)$ calculates the rate at which the jerk parameter changes. These parameters are defined as follows:
\begin{equation}{\label{109}}
	\mathit{j}=\frac{1}{aH^{3}}\left(\frac{d^{3}a}{dt^{3}}\right),\ \ \mathit{s}=\frac{1}{aH^{4}}\left(\frac{d^{4}a}{dt^{4}}\right)
\end{equation}
In the present analysis, for $j$ and $s$, equation (\ref{109}) can be rewritten in terms of $z$ as \cite{wang2009probing}.\\
\begin{equation}{\label{1099}}
	j(z)=q(z)\left(1+2q(z)\right)+\frac{dq}{dz}(1+z), \  \ \  \   s(z)=-\frac{dj}{dz}(1+z)-j(z)\left(2+3q(z)\right).
\end{equation}
For Model I, based on the constrained values from the CC data set, the snap parameter is $s_0=-2.12$ and the jerk parameter $ j_0=1.14$. However, from the joint estimates, the jerk parameter value is $ j_0=1.25$ and snap parameter value is $s_0=-2.03$. Similarly, for Model II, the value of the jerk parameter is $j_0= 2.35$ (for CC data set) and $j_0= 1.26$ (for joint data set). In the Tables (\ref{table:1}) and (\ref{table:2}), we summarize the present day values of the cosmological parameters. For Model I, the value of jerk parameter exhibits a decreasing trend from early to late times, eventually approaching to $1$. This behavior suggests that these models differ from $\Lambda$CDM in the early universe but the Model I would become similar to $\Lambda$CDM at later times.  

\subsection{The physical behavior of models}
\label{sec:6.2}
\begin{figure}[!htb]
\captionsetup{skip=0.3\baselineskip,size=footnotesize}
   \begin{minipage}{0.5\textwidth}
     \centering
     \includegraphics[width=8.0 cm,height=7.0cm]{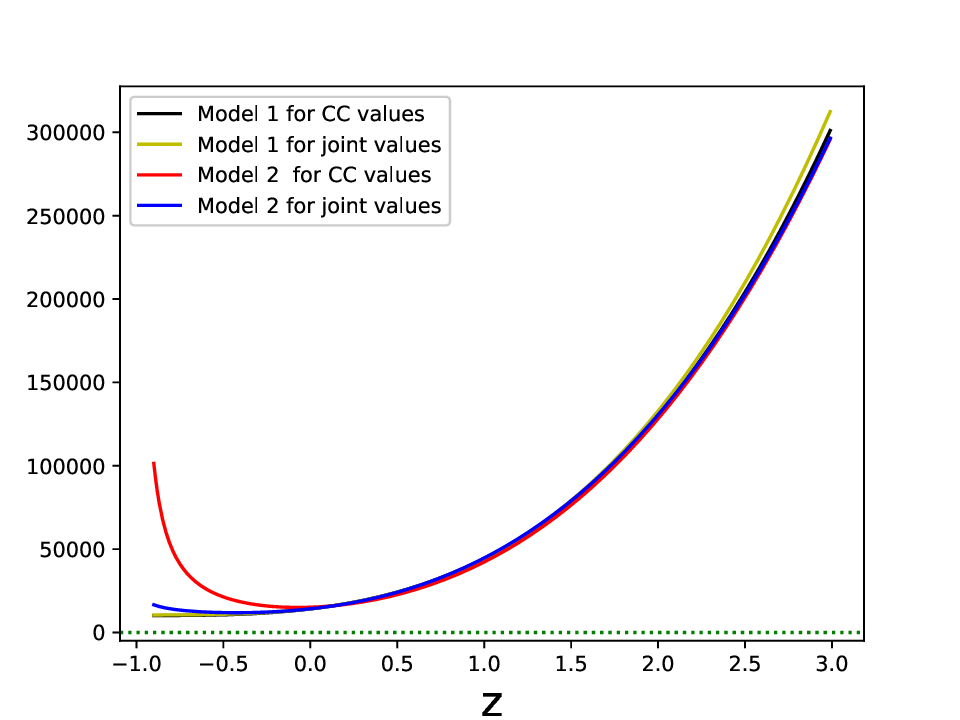}
\caption{Total energy density versus $\mathit{z}$ }
\label{fig:9}
    \end{minipage}\hfill
   \begin{minipage}{0.5\textwidth}
     \centering
     \includegraphics[width=8.0cm,height=7.0cm]{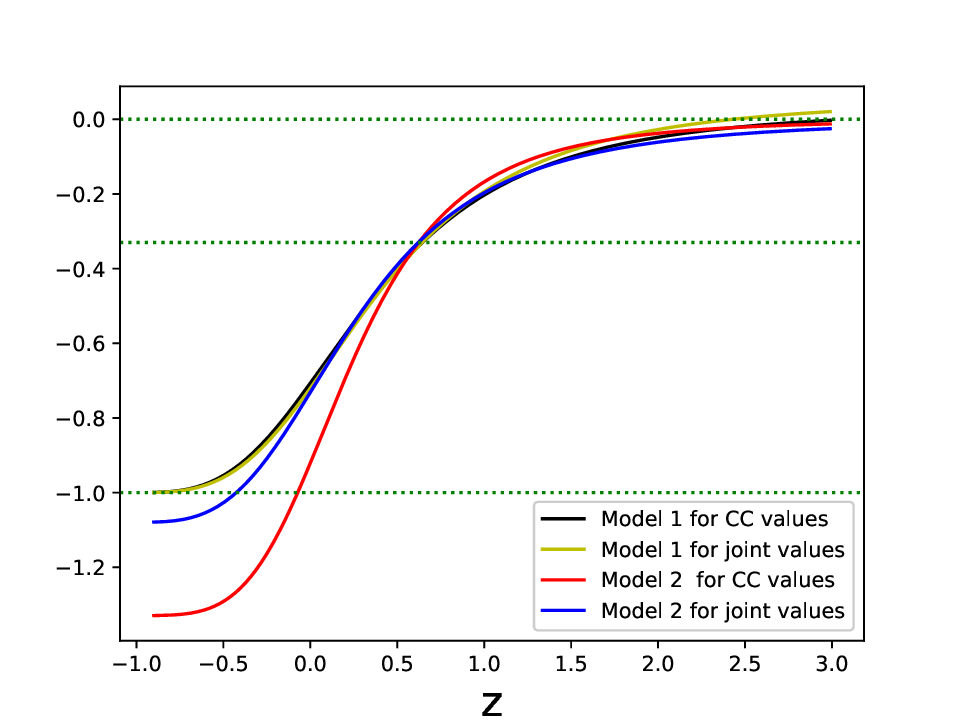}
  \caption{$\omega_e$ versus $\mathit{z}$ }
\label{fig:13}
   \end{minipage}
\end{figure}

We highlight the physical behavior exhibited by the quantities such as the energy density and pressure. For the model parameters' constrained values, the energy density has remained positive throughout the expansion history while the pressure may have undergone a recent transition to negative values. We calculate the energy density and pressure by utilizing field equations and the Eqs. (\ref{301}) and (\ref{302}). In model I, the dark fluid having energy density $\rho_d$ will act as the total energy density of universe in the context of General relativity. And, the Model II possess the matter and the dark energy (due to dark fluid) will have the total energy density $\rho_e$.\\
The evolution of energy densities are depicted in Figs. (\ref{fig:9}) for Model I and II. The energy densities in these models will be increasing with $z$ (decreases with cosmic time $(t)$), remaining positive throughout expansion for both the models. In other words, the energy density will be decreasing from past to future in Model I. And, thus the energy density will have positive values and it rule out the existence of finite-time future singularities. In Model II, the energy density possess phantom nature. During the early universe, at large $z$, the pressure commences with the positive values. In the later epoch along with the present epoch, the pressure becomes negative. It is observed from the graphical analysis that the non-metricity scalar will be responsible for the negative pressure necessary for the accelerated universe expansion in this late epoch. \\ 
The EoS parameter may offer insights on the dominating fluid during the cosmic expansion of the universe and thus helps to determine the dynamics of the Universe. In the cosmological models, it may be commonly used to identify the nature of dark energy. The EoS parameter can be defined using the relationship between pressure $(p)$ and energy density $(\rho)$. It is mathematically expressed as $(\omega=\frac{p}{\rho})$. The cosmic characteristics, including the cold dark matter or dust dominated phase will have $\omega = 0$ and the radiation-dominated period will have $\omega=\frac{1}{3}$. The transition into an accelerated cosmic era happens for the equation of state parameter satisfying  $\omega < -\frac{1}{3}$. The cosmic acceleration period may have three regimes: the quintessence-dominated regime for $-1 < \omega < -\frac{1}{3}$, the cosmological constant-like state for $\omega = -1$ and phantom-dominated regime $\omega < -1$. In the Fig. (\ref{fig:13}), we provide the graphical representation of the EoS parameter evolution in the considered models.
In the Model I, the EoS parameter (defined as $\omega_e=p_d/\rho_d$) values are estimated as $\omega_{e}=-0.706$ and $\omega_{e}=-0.720$ for CC and joint dataset respectively at $z=0$. The Model II possess matter and dark energy, as a consequence, we define $\omega_e=p_e/\rho_e$, where the total energy density and pressure are denoted by $\rho_e$ and $p_e$ respectively. In the Model II, the effective EoS parameter values are obtained $\omega_{e} = -0.921$ (for CC dataset) and $\omega_{e} = -0.73$ (for joint dataset) at $z=0$. The Fig. (\ref{fig:13}) is revealing that the effective EoS parameter may correspond to the quintessence kind of dark energy in these models at present times. In future, as $z\rightarrow -1$, the cosmological constant-like behavior may be visualized in the Model I. On the other hand, in the Model II, the phantom kind of dark energy may evolve during the cosmological evolution. In the present era, the dark energy is a dynamical one which is varying with time in these models. \\
These models possess the past of matter-dominated expansion which are confirmed by Fig. (\ref{fig:13}) having $\omega=0$ during early times. The cosmic evolution is under the domination of quintessence kind of dark energy having $-1 < \omega < -\frac{1}{3}$ in the present epoch. However, the future evolution in both of these models may be different. In the model I, the dark fluid is in charge of the matter dominated expansion and the dark energy dominated expansion. In this sense, the unified scenario has been confirmed in the affine EoS model of the General relativity. On the other hand, when the cold dark matter has been considered with the dark fluid satisfying affine EoS, the $f(Q)$ gravity model yields the identical matter evolution as of the $\Lambda$CDM model but the dark energy behavior will be clearly different. In the Model I, the fluid with affine EoS unifies the decelerated and accelerated expansion. While in the Model II with cold dark matter and dark fluid having affine EoS, the matter dominated era may be the same but phantom nature of dark energy may not be ruled out in this model. 
\subsection{Age of the Universe}
\label{sec:6.5}
The age of the universe is an important criterion to test the viability of the model as compared to the observable universe. The cosmic age $t(z)$ may be expressed in terms of the $z$ as ~\cite{tong2009cosmic}
\begin{equation}{\label{34}}
t(z)= \int_{z}^{\infty} \frac{dz'}{(1+z')H(z')}. 
\end{equation}
The present age $(t_{0})$ at $(z=0)$ has been calculated numerically using the integral consisting of the Hubble parameters $H(z)$ provided in equations (\ref{301}) and (\ref{302}). \\
In the  Model I, the universe's current  age is $t_{0}$ (at $z=0)=13.51^{+0.08}_{-0.37}$ Gyr for CC data sets and $t_{0}=13.41^{+0.65}_{-1.04}$ Gyr for joint data sets, both of which are close with the current age values of the $\Lambda$CDM model (having $t_0=13.79$ Gyr) obtained from Planck results \cite{2020A&A...641A...6P}. \\
In model II, the universe present age value is $t_{0} = 13.76$ Gyr and $t_{0} = 13.59$ Gyr for CC and joint estimates respectively, both of which are again close with the current age values of the $\Lambda$CDM model mentioned above. 
\subsection{Model I as the $f(Q)$ gravity model?}
\label{sec:3.1}
The Hubble parameter may be visualized as the solution of field equations and thus may be used for the identification of cosmic dynamics in a model. In the General relativity model with dark fluid (\ref{eqr2}), the expansion rate is described by Hubble parameter (\ref{301}). Without imposing any particular form of $f(Q)$ in modified gravity model, we probe whether this Hubble parameter may be obtained in $f(Q)$ gravity or not? In other words, we search for the functional form of $f(Q)$ in the $f(Q)$ gravity, where $\rho_d$ of Eq. (\ref{14}) is same as that of the energy density of fluid satisfying (\ref{eqr2}). \\
In order to search for the equivalence of $f(Q)$ gravity with the framework having Hubble parameter (\ref{301}) governed by the dark fluid (\ref{eqr1}) only, we solve the Eq. (\ref{14}) with $Q=6H^2$ and $k^2\rho_d=3H^2$. It would lead to $F(Q) = -Q + C\sqrt{Q}$, where $C$ is an integration constant. The form of $f(Q)$ function may be written as 
\begin{equation}
	f(Q)=C\sqrt{Q}.
\end{equation}
It is well known that $f(Q)=C\sqrt{Q}$ model is a total derivative in Friedmann-Lemaitre-Robertson-Walker cosmological framework and thus FLRW field equations of $f(Q)$ gravity will vanish identically\cite{dimakis2022flrw,dimakis2021quantum,
capozziello2022comparing,snsdo2025fq}. It means that the solution (\ref{301}) will be of the General Relativity but not of the $f(Q)$ gravity.

\section{Conclusions}
\label{sec:7}
We probe the possibility to describe the unified scenario of universe expansion with the affine equation of state (EoS) in the General relativity and $f(Q)$ gravity. The late-time accelerated expansion of the Universe has been investigated with an affine EoS satisfying $p_d=\alpha\rho_d-\rho_0$. We solve the continuity equation for the affine EoS in General relativity and show that the Hubble parameter of model may unify the decelerating-accelerating universe expansion. We further obtain the observational constraints on model parameters of this model. Without imposing any particular form of $f(Q)$ in context of $f(Q)$ gravity, we probe whether the Hubble parameter (\ref{301}) of Model I may be obtained in $f(Q)$ gravity or not? It leads us to the functional form $f(Q)=C\sqrt{Q}$, (where $C$ is constant) in context of $f(Q)$ gravity, which means that this solution (\ref{301}) will only be of the General relativity due to the fact that $f(Q)=C\sqrt{Q}$ model is total derivative in the FLRW context and thus $f(Q)$ field equations will vanish identically. We further study the $f(Q)$ model composed of cold dark matter and the dark fluid satisfying the affine EoS. This model governed by $H(z)$ (\ref{302}) describes the phantom evolution of dark energy, subjected to the observational constraints. A detailed descriptive differences between these models have been studied by using different cosmological identifiers. For this model, we numerically solved the differential equation (\ref{1011}) with the initial condition $f(0)=6H_{0}^{2}(2+\Omega_{d0})$. The evolution of the functions $F(Q)$ and $f(Q)$ with redshift is presented in figures $(\ref{fig:109})$ and $(\ref{fig:113})$ respectively.   Figure $(\ref{fig:109})$ and $(\ref{fig:113})$ clearly shows that both $F$ and $f$ are decreasing functions as the universe transitions from high redshift to the current epoch. The higher values of these functions in the early epoch (large z) correspond to stronger non-metricity effects dominating the universe’s dynamics in the past. As the redshift decreases toward the present time, value of $F(Q)$ and $f(Q)$ gradually approach smaller, which consistent with observational evidence of a transition from decelerated to accelerated expansion.\\
We use the observational data of the CC and supernovae type Ia Pantheon data for the MCMC analysis based on the Hubble parameters (\ref{301}) and (\ref{302}) of the derived models. A summary of the constrained values extracted from the MCMC application in the models are provided in Tables (\ref{table:1}) and (\ref{table:2}). \\
The evolution dynamics in the present study has been examined by study of the cosmological parameters. The behavior of the deceleration parameter has been investigated. The evolution and dominant cosmic components during various epochs are reflected in the deceleration parameter behaviors of both models. The universe’s phase recently transitioned from decelerated to accelerated as indicated by the deceleration parameter’s evolution curve. In Model I, the estimated present deceleration parameter values are $q_{0}=-0.56$ and $q_{0}=-0.58$ for the CC and joint analyses respectively. For Model II, we obtain the deceleration parameter values at present as $q_{0}=-0.881$ and $q_{0}=-0.597$ based on the CC and joint data respectively. A negative deceleration parameter at $z=0$ demonstrates that cosmic expansion is accelerating for both models. At late-times, the negative pressure introduced by the dark fluid may be responsible for the accelerated universe expansion.\\
In the considered models, the physical behavior has been elaborated on the basis of energy density, pressure and EoS parameters for the model parameters' constrained values. The energy density remains positive whereas pressure exhibits a change from positive to negative values. It is observed in the models that the cosmic accelerated expansion is caused by this negative pressure available in the late era of the universe expansion. The universe in Model I may approach to the $\Lambda$CDM limits in future as $z\rightarrow -1$. The varying dark energy evolution may correspond to the cosmological constant-like behavior in the asymptotic era having $z\rightarrow -1$. While the energy density of the universe in Model II will increase in future due to phantom-like characteristics. The parameter constraints subjected to the CC and Pantheon data suggest that Model I is generally following the classical stability criterion for broader region of parameter space within $1\sigma$ limits. On the other hand, the Model II describing phantom evolution of dark energy is classically unstable within $1\sigma$ limits. It is an important distinction as far as physical acceptability of these models are concerned. \\
In particular, for the model I, at $z = 0$, $\omega_{e} = -0.706$  and $\omega_{e} = -0.720$ for CC and joint data respectively. These values at the present time suggest the existence of a quintessence-type dark energy in the model. Similarly, for the Model II, the EoS parameter values are obtained $\omega_{e} = -0.921$ (for CC dataset) and $\omega_{e} = -0.731$ (for joint dataset) at $z=0$. The graphical behavior reveals that the EoS parameter for dark energy exhibits quintessence kind of dark energy at present times. In summary, the affine EoS model is consistent with the observations of cosmic acceleration. The violation of null energy condition in the Model II may not prohibit from the existence of finite-time future singularity.\\
For the $\Lambda$CDM model, the cosmological parameters are $q=-1$, $\omega = -1 $, $j=1$, Snap $(s)=1$ during the asymptotic era as $z\rightarrow -1$. In other words, in the de-Sitter model of cosmic expansion, where $H$ remains constant and $q=-1$, $j=1$ and $s=1$. It indicates that the universe is undergoing exponential acceleration with the rate of change of both acceleration and jerk increasing. In this context, for the Model I (Eq. \ref{301}), the dark energy equation of state (EoS) parameter ($ \omega_{e})$ at present is near $-0.7 $ and the deceleration parameter is approximately $-0.56$ (for CC data) and $-0.58$ (for joint data). In this scenario, the universe's acceleration is attributed to dynamical dark energy and the jerk parameter is close to $\Lambda$CDM limit $(j=1)$ while the snap parameter is negative for the median values of model parameters. The Tables (\ref{table:1}) and (\ref{table:2}) highlight a summary of the results on cosmographic parameters. It is worthwhile to mention that these quantities will eventually align with the $\Lambda$CDM limits in the distant future. However, at present times, the cosmic dynamics of these models are differing from the $\Lambda$CDM model. \\
The age of the universe is also estimated in the context of present $f(Q)$ gravity models. For model I, the value is $t_{0} = 13.51$ Gyr  and $t_{0} = 13.41$ Gyr for CC and joint estimates respectively. For model II, the universe's present age value is $t_{0} = 13.76 $ Gyr (for CC data) and $t_{0} = 13.59$ Gyr (for joint data).\\
In summary, we show that the observationally compatible unified scenario may be realized in the General relativity and $f(Q)$ gravity frameworks with the affine EoS. The Hubble parameters of the considered models may explain the matter dominated era of the past along with the dark energy dominated era of the observable universe. A descriptive difference between two class of models have illustrated that the present day matter density of the considered models are approximately the same. However, in future, the Model I possess quintessence kind of dark energy and is classically stable for broader region of parameter space while Model II possess phantom kind of dark energy and is classically unstable. Additionally, the dynamical dark energy in Model I is of quintessence kind with $\Lambda$CDM limit in future while the dynamical dark energy of Model II may have the phantom nature in future. 
\small
\section*{Acknowledgments}
We are grateful to the reviewers for comments which have been very helpful to modify the paper. 
GPS and AS are thankful to the Inter-University Centre for Astronomy and Astrophysics (IUCAA), Pune, India for support under Visiting Associateship program. 


\begin{thebibliography}{99}
\bibitem{1998AJ....116.1009R} A. G. Riess, et al., Observational Evidence from Supernovae for an Accelerating Universe and a Cosmological Constant, Astron. J. \textbf{116}, 1009 (1998). 

\bibitem{1999ApJ...517..565P} S. Perlmutter et al., Measurements of $\Omega$ and $\Lambda$ from 42 High-Redshift Supernovae, Astrophys. J. \textbf{517}, 565 (1999). 

\bibitem{knop2003new} R. A. Knop et al., New constraints on $\Omega_{M}$, $\Omega_{\Lambda}$ and $w$ from an independent set of $11$ high-redshift supernovae observed with the Hubble Space Telescope, Astrophys. J. \textbf{598}, 102 (2003). 

\bibitem{tonry2003cosmological} J. L. Tonry et al., Cosmological results from high $z$ supernovae, Astrophys. J.  \textbf{594}, 1 (2003).  

\bibitem{2020A&A...641A...6P} N. Aghanim et al., Planck 2018 results VI Cosmological parameters, Astron. Astrophys. \textbf{641}, A6 (2020). 

\bibitem{padmanabhan2003cosmological} T. Padmanabhan, Cosmological constant the weight of the vacuum, Phys. Rep. \textbf{380}, 235-320 (2003).

\bibitem{carroll2001cosmological} S. M. Carroll, The cosmological constant, Living Rev. Relativ. \textbf{4}, 1-56 (2001). 

\bibitem{sahni2000case} V. Sahni, A. Starobinsky, The case for a positive cosmological $\Lambda$ term, Int.
J. Mod. Phys. D, \textbf{9}, 373-443 (2000).   

\bibitem{tsujikawa2010modified} S. Tsujikawa, Modified gravity models of dark energy, Lectures on cosmology: Accelerated Expansion of the Universe, 99-45 (2010).    

\bibitem{nojiri2011unified} S. Nojiri, S. D. Odintsov, Unified cosmic history in modified gravity: from $f(R)$ theory to Lorentz non-invariant models, Phys. Rep. \textbf{505}, 59--144 (2011). 

\bibitem{nojiri2017modified} S. Nojiri, S. D. Odintsov, V. K. Oikonomou, Modified gravity theories on a nutshell: Inflation, bounce and late-time evolution, Phys. Rep. \textbf{692}, 1-104 (2017).

\bibitem{de2023finite} J. de Haro, S. Nojiri, S. D. Odintsov, V. K. Oikonomou, S. Pan, Finite-time cosmological singularities and the possible fate of the Universe, Phys. Rep., \textbf{1034} 1-114 (2023).     

\bibitem{di2021realm} E. Di Valentino et al., In the realm of the Hubble tension a review of solutions, Class. Quantum
Grav., \textbf{38}, 153001 (2021).     

\bibitem{heisenberg2024review} L. Heisenberg, Review on $f(Q)$ gravity, Phys. Rep., \textbf{1066}, 1-78 (2024).    

\bibitem{weinberg1989cosmological} S. Weinberg, The cosmological constant problem, Rev. Mod. Phys. \textbf{61}, 1 (1989). 

\bibitem{buchdahl1970non} H. A. Buchdahl, Non-linear Lagrangians and cosmological theory, Mon. Not. R. Astronom. Soc. \textbf{150}, 1-8 (1970). 

\bibitem{harko2011f} T. Harko, F. S. N. Lobo, S. Nojiri, S. D. Odintsov, $f(R, T)$ gravity, Phys. Rev. D, \textbf{84}, 024020 (2011).    

\bibitem{elizalde2010lambdacdm} E. Elizalde, R. Myrzakulov, V. V. Obukhov, D. Saez-Gomez, $\Lambda$CDM epoch reconstruction from $F(R, G)$ and modified Gauss Bonnet gravities, Class. Quantum
Grav.,  \textbf{27}, 095007 (2010).      

\bibitem{bamba2010finite} K. Bamba, S. D. Odintsov, L. Sebastiani, S. Zerbini, Finite time future singularities in modified Gauss Bonnet and $F(R, G)$ gravity and singularity avoidance, Eur. Phys. J. C,  \textbf{67}, 295-310 (2010).   

\bibitem{harko2010f} T. Harko, F. S. N. Lobo, $f(R, L_{m})$ gravity, Eur. Phys. J. C, \textbf{70}, 373-379 (2010).     

\bibitem{chaubey2016general} R. Chaubey, A. K. Shukla, R. Raushan, T. Singh, The general class of Bianchi cosmological models in $f(R, T)$ gravity with dark energy in viscous cosmology, Indian J. Phys., \textbf{90}, 233-242 (2016).     

\bibitem{fqt} Y. Xu, G. Li, T. Harko, S. D. Liang, $f(Q, T)$ gravity, Eur. Phys. J. C, \textbf{79}, 708 (2019).    

\bibitem{extcos} S. Capozziello, R. D'Agostino, O. Luongo, Extended gravity cosmography, Int. J. Mod. Phys. D \textbf{28}, 1930016 (2019). 	

\bibitem{singh2020study} G. P. Singh, A. R. Lalke, N. Hulke, Study of particle creation with quadratic equation of state in higher derivative theory, Braz. J. Phys. \textbf{50}, 725-743 (2020). 

\bibitem{as2022lyra} A. Singh, Qualitative study of Lyra cosmologies with spatial curvature, Chinese J. Phys., \textbf{79}, 481-489 (2022).

\bibitem{Garg2023cfn} R. Garg, G. P. Singh, A. R. Lalke, S. Ray, Cosmological model with linear equation of state parameter in $f(R, L_{m})$ gravity, Phys. Lett. A, \textbf{525}, 129937 (2024).   

\bibitem{lalke2023late} A. R. Lalke, G. P. Singh, A. Singh, Late-time acceleration from ekpyrotic bounce in $f(Q, T)$ gravity, Int. J. Geom. Methods Mod. Phys., \textbf{20}, 2350131 (2023).    

\bibitem{as2024lyra} A. Singh, Lyra cosmologies with the dynamical system perspective, Phys. Scr., \textbf{99}, 045011 (2024).  

\bibitem{capozziello2023role} S. Capozziello, V. De Falco, C. Ferrara, The role of the boundary term in $f(Q, B)$ symmetric teleparallel gravity, Eur. Phys. J. C, \textbf{83}, 915 (2023).       

\bibitem{singh2024abcde123} G. P. Singh, R. Garg, A. Singh, A generalized $\Lambda$CDM model with parameterized Hubble parameter in particle creation, viscous and $f(R)$ model framework, Int. J. Geom. Methods Mod. Phys., \textbf{22}, 2550111 (2025).  

\bibitem{mgrav1} A. Singh, Dynamical systems of modified Gauss–Bonnet gravity: cosmological implications, Eur. Phys. J. C, \textbf{85}, 24 (2025).  

\bibitem{jimenez2018coincident} J. B. Jimenez, L. Heisenberg, T. Koivisto, Coincident general relativity, Phys. Rev. D, \textbf{98}, 044048 (2018).     

\bibitem{yang20212021} W. Yang, S. Pan, E. Di Valentino, O. Mena, A. Melchiorri, 2021-$H_0$ odyssey: closed, phantom and interacting dark energy cosmologies, JCAP,  \textbf{2021}, 008 (2021). 

\bibitem{lazkoz2019observational} R. Lazkoz, F. S. N.  Lobo, M. Ortiz Banos, V. Salzano, Observational constraints of $f(Q)$ gravity, Phys. Rev. D, \textbf{100}, 104027 (2019).     

\bibitem{esposito2022reconstructing} F. Esposito, S. Carloni, R. Cianci, S. Vignolo, Reconstructing isotropic and anisotropic $f(Q)$ cosmologies, Phys. Rev. D, \textbf{105}, 084061 (2022).        

\bibitem{harko2018coupling} T. Harko, T. S. Koivisto, F. S. N. Lobo, G. J. Olmo, D. Rubiera Garcia, Coupling matter in modified $Q$ gravity, Phys. Rev. D, \textbf{98} 084043 (2018).     

\bibitem{maurya2023transit} D. C. Maurya, A. Dixit, A. Pradhan, Transit string dark energy models in $f(Q)$ gravity, Int. J. Geom. Methods Mod. Phys., \textbf{20} 2350134 (2023).    

\bibitem{pradhan2022quintessence} A. Pradhan, A. Dixit, D. C. Maurya, Quintessence Behavior of an Anisotropic Bulk Viscous Cosmological Model in Modified $f(Q)$ Gravity, Symmetry \textbf{14}, 2630 (2022).

\bibitem{capozziello2024preserving} S. Capozziello, A. Lapponi, O. Luongo, S. Mancini, Preserving quantum information in $f(Q)$ non-metric gravity cosmology, Eur. Phys. J. C, \textbf{84}, 1081 (2024).      

\bibitem{nojiri2024well} S. Nojiri, S. D. Odintsov, Well-defined $f(Q)$ gravity, reconstruction of FLRW spacetime and unification of inflation with dark energy epoch, Phys. Dark Uni., \textbf{45}, 101538 (2024).   

\bibitem{hu2023nonpropagating} K. Hu, M. Yamakoshi, T. Katsuragawa, S. Nojiri, T. Qiu, Nonpropagating ghost in covariant $f(Q)$ gravity, Phys. Rev. D, \textbf{108}, 124030 (2023).       

\bibitem{capozziello2022model} S. Capozziello, R. D'Agostino, Model independent reconstruction of $f(Q)$ non metric gravity, Phys. Lett. B, \textbf{832}, 137229 (2022).  

\bibitem{khyllep2021cosmological} W. Khyllep, A. Andronikos, J. Dutta, Cosmological solutions and growth index of matter perturbations in $f(Q)$ gravity, Phys. Rev. D, \textbf{103}, 103521 (2021).      

\bibitem{dimakis2021quantum} N. Dimakis, A. Paliathanasis, T. Christodoulakis, Quantum cosmology in $f(Q)$ theory, Class. Quantum
Grav., \textbf{38}, 225003 (2021).  

\bibitem{dimakis2022flrw} N. Dimakis, A. Paliathanasis, M. Roumeliotis, T. Christodoulakis, FLRW solutions in $f(Q)$ theory: The effect of using different connections, Phys. Rev. D, \textbf{106}, 043509 (2022). 

\bibitem{heisenberg2024cosmological} L. Heisenberg, M. Hohmann, S. Kuhn, Cosmological teleparallel perturbations, JCAP, \textbf{2024}, 063 (2024). 

\bibitem{capozziello2022comparing} S. Capozziello, V. De Falco, C. Ferrara,  Comparing equivalent gravities: common features and differences, Eur. Phys. J. C, \textbf{82}, 865 (2022).  

\bibitem{koussour2023square} M. Koussour et al., Square-root parametrization of dark energy in $f(Q)$ cosmology, Commun. Theor. Phys., \textbf{75}, 125403 (2023).    

\bibitem{rani2025physical} S. Rani, M. Adeel, M. Z. Gul, A. Jawad, S. Shaymatov, Physical viability of $f(Q)$ gravity corrected charged anisotropic solutions, Phys. Dark Uni., \textbf{47}, 101754 (2025). 

\bibitem{sarmah2024dynamical} P. Sarmah, U. D. Goswami, Dynamical system analysis of LRS-BI Universe with $f(Q)$ gravity theory, Phys. Dark Uni., \textbf{46}, 101556 (2024).  

\bibitem{nashed2024general} G. G. L. Nashed, S. Nojiri, General geometry realized by four-scalar model and application to $f(Q)$ gravity, Phys. Dark Uni., \textbf{46}, 101655 (2024).    

\bibitem{capozziello2024gravitational} S. Capozziello, M. Capriolo, Gravitational waves in $f(Q)$ non-metric gravity without gauge fixing, Phys. Dark Uni., \textbf{45}, 101548 (2024).    

\bibitem{parsaei2022wormhole} F. Parsaei, S. Rastgoo, P. K. Sahoo, Wormhole in $f(Q)$ gravity, Eur. Phys. J. Plus, \textbf{137}, 1083 (2022).      

\bibitem{narawade2025stable} S. A. Narawade, S. V. Lohakare, B. Mishra, Stable $f(Q)$ gravity model through non-trivial connection, Ann. Phys., \textbf{474}, 169913 (2025).

\bibitem{Sahlu1} S. Sahlu, A. de la Cruz-Dombriz, A. Abebe, Structure growth in $f(Q)$ cosmology, Mon. Not. R. Astron. Soc., \textbf{539}, 690-703 (2025). 

\bibitem{Sahlu2} S. Sahlu, R. T. Hough, A. Abebe, A. de la Cruz-Dombriz, Constraining viscous-fluid models in $f(Q)$ gravity with data, Eur. Phys. J. C, \textbf{85}, 746 (2025).  

\bibitem{singha2025} R. Singha, A. Singh, Observationally compatible cosmological scenarios in $f(Q)$ gravity with Lagrangian reconstruction, Gravit. Cosmol.  \textbf{31}, 260-269 (2025).

\bibitem{garg2024observational} R. Garg, G. P. Singh, A. Singh, Late-time dynamics of dark energy EoS in symmetric teleparallel gravity, Int. J. Mod. Phys. A, \textbf{40}, 2550099 (2025). 

\bibitem{fluid1} E. Babichev, V. Dokuchaev, Y. Eroshenko, Dark energy cosmology with generalized linear equation of state, Class. Quantum Grav., \textbf{22}, 143 (2005). 

\bibitem{fluid8} T. Chiba, N. Sugiyama, T. Nakamura, Cosmology with x-matter, Mon. Not. R. Astron. Soc., \textbf{289}, L5-L9 (1997).      

\bibitem{fluid9} E. Babichev, V. Dokuchaev, Y. Eroshenko, Black Hole Mass Decreasing due to Phantom Energy Accretion, Phys. Rev. Lett., \textbf{93}, 021102 (2004).

\bibitem{fluid6} H. Stefancic, Expansion around the vacuum equation of state: Sudden future singularities$<?$ format$?>$ and asymptotic behavior, Phys. Rev. D \textbf{71}, 084024 (2005).    

\bibitem{fluid3} K. Ananda, M. Bruni, Cosmological dynamics and dark energy with a nonlinear equation of state: A quadratic model, Phys. Rev. D, \textbf{74}, 023523 (2006).    

\bibitem{fluid2} A. Balbi, M. Bruni, C. Quercellini, $\Lambda\alpha$DM: Observational constraints on unified dark matter with constant speed of sound, Phys. Rev. D, \textbf{76}, 103519 (2007).      

\bibitem{fluid4} T. Singh, R. Chaubey, Bianchi Type-I Universe with wet dark fluid, Pramana J. Phys., \textbf{71}, 447–458 (2008).    

\bibitem{fluid5} R. Chaubey, Bianchi type-V universe with wet dark fluid, Astrophys. Space Sci., \textbf{321}, 241-246 (2009).   

\bibitem{fluid7} G. S. Khadekar, Brane Kantowski-Sachs universe with linear equation of state and a future singularity, Gravit. Cosmol., \textbf{21}, 334–339 (2015).

\bibitem{fluid11} G. P. Singh, N. Hulke, A. Singh, Thermodynamical and observational aspects of cosmological model with linear equation of state, Int. J. Geom. Methods Mod. Phys., \textbf{15}, 1850129 (2018).   

\bibitem{fluid13} A. Singh, R. Raushan, R. Chaubey, Qualitative aspects of Rastall gravity with barotropic fluid, Can. J. Phys., \textbf{99}, 1073-1081 (2021).    

\bibitem{fluid14} A. Singh, G. P. Singh, A. Pradhan, Cosmic dynamics and qualitative study of Rastall model with spatial curvature, Int. J. Mod. Phys. A, \textbf{37}, 2250104 (2022).      

\bibitem{fluid15} A. Singh, Homogeneous and anisotropic cosmologies with affine EoS: a dynamical system perspective, Eur. Phys. J. C \textbf{83}, 696 (2023). 

\bibitem{fluid16} A. Singh, S. Krishnannair, Affine EoS cosmologies: Observational and dynamical system constraints, Astron. Comput. \textbf{47}, 100827 (2024). 

\bibitem{cs1} P. J. E. Peebles, B. Ratra, The cosmological constant and dark energy, Rev. Mod. Phys. \textbf{75}, 559-606 (2003).      

\bibitem{cs2} G. F. R. Ellis, R. Maartens, M. A. H. MacCallum, Causality and the speed of sound, Gen. Relativ. Grav., \textbf{39}, 1651–1660 (2007).      

\bibitem{PhysRevLett.80.1582} R. R. Caldwell, R. Dave, P. J. Steinhardt, Cosmological Imprint of an Energy Component with General Equation of State, Phys. Rev. Lett., \textbf{80}, 1582-1585 (1998).     

\bibitem{caldwell2002phantom} R. R. Caldwell, A phantom menace? Cosmological consequences of a dark energy component with super-negative equation of state, Phys. Lett. B, \textbf{545}, 23-29 (2002).   

\bibitem{cai2010quintom} Y. F. Cai, E. N. Saridakis, M. R. Setare, J. Q. Xia, Quintom cosmology: theoretical implications and observations, Phys. Rep., \textbf{493}, 1-60 (2010).  

\bibitem{jimenez2020cosmology} J. B. Jimenez, L. Heisenberg, T. Koivisto, S. Pekar, Cosmology in $f(Q)$ geometry, Phys. Rev. D, \textbf{101} 103507 (2020).        

\bibitem{foreman2013emcee} D. Foreman-Mackey, D. W. Hogg, D. Lang, J. Goodman, emcee: the MCMC hammer, Pub. Astron. Soc. Pac. \textbf{125}, 306 (2013). 

\bibitem{vagnozzi2021eppur} S. Vagnozzi, A. Loeb, M. Moresco, Eppure piatto? The cosmic chronometers take on spatial curvature and cosmic concordance, Astrophys. J., \textbf{908}, 84 (2021). 

\bibitem{GSSharov} G. S. Sharov, V. O. Vasiliev, How predictions of cosmological models depend on Hubble parameter data sets, Math. Model. Geom., \textbf{6}, 1 (2018).     

\bibitem{jimenez2002constraining} R. Jimenez, A. Loeb, Constraining cosmological parameters based on relative galaxy ages, Astrophys. J., \textbf{573}, 37 (2002).    
  
\bibitem{lalke2024cosmic} A. R. Lalke, G. P. Singh, A. Singh, Cosmic dynamics with late-time constraints on the parametric deceleration parameter model, Eur. Phys. J. Plus \textbf{139}, 288 (2024).

\bibitem{mandal2024late} S. Mandal, A. Singh, R. Chaubey, Late-time constraints on barotropic fluid cosmology, Phys. Lett. A,\textbf{519} 129714 (2024).    

\bibitem{aspdu} A. Singh, S. Mandal, R. Chaubey, R. Raushan, Observational constraints on the expansion scalar and shear relation in the Locally rotationally symmetric Bianchi I model, Phys. Dark Uni., \textbf{47} 101798 (2025).   

\bibitem{sma2026} S. Mandal, Observational constraints on new class dark energy parameterized EoS in Bianchi type-I universe, JHEAP, \textbf{52},  100564 (2026).

\bibitem{scolnic2018complete} D. M. Scolnic et al., The complete light-curve sample of spectroscopically confirmed SNe Ia from Pan-STARRS1 and cosmological constraints from the combined pantheon sample, Astrophys. J., \textbf{859}, 101 (2018).   

\bibitem{ellis2012relativistic} G. F. R. Ellis, R. Maartens, M. A. H. MacCallum, Relativistic cosmology, (2012) Cambridge University Press.  

\bibitem{asvesta2022observational} K. Asvesta, L. Kazantzidis, L. Perivolaropoulos, C. G. Tsagas, Observational constraints on the deceleration parameter in a tilted universe, Mon. Not. R. Astron. Soc.,  \textbf{513}, 2394-2406 (2022).       

\bibitem{weinberg2008cosmology} S. Weinberg, Cosmology (Oxford University Press, UK, 2008). 

\bibitem{mukherjee2016parametric} A. Mukherjee, N. Banerjee, Parametric reconstruction of the cosmological jerk from diverse observational data sets, Phys. Rev. D \textbf{93} 043002 (2016).

\bibitem{mandal2022epjp} S. Mandal, A. Singh, R. Chaubey, Observational constraints and cosmological implications of NLE model with variable $G$, Eur. Phys. J. Plus \textbf{137}, 1246 (2022).

\bibitem{SINGH2024865} A. Singh, Qualitative study of anisotropic cosmologies with inhomogeneous equation of state, Chinese J. Phys. \textbf{88}, 865-878 (2024)

\bibitem{asgrg2} A. Singh, Role of dynamical vacuum energy in the closed universe: implications for bouncing scenario, Gen. Relativ. Gravit. \textbf{56}, 138 (2024). 

\bibitem{wang2009probing} F. Y. Wang, Z. G. Dai, S. Qi, Probing the cosmographic parameters to distinguish between dark energy and modified gravity models, Astron. $\&$ Astrophys, \textbf{507}, 53-59 (2009).    

\bibitem{tong2009cosmic} M. L. Tong, Y. Zhang, Cosmic age, statefinder, and Om diagnostics in the decaying vacuum cosmology, Phys. Rev. D, \textbf{80}, 023503, (2009). 

\bibitem{snsdo2025fq} S. Nojiri, S. D. Odintsov, Well-defined $f(Q)$ gravity, reconstruction of FLRW spacetime and unification of inflation with dark energy epoch, Phys. Dark Uni., \textbf{45}, 101538, (2024).

\end{thebibliography}


\end{document}